\documentclass[aps,prb,twocolumn,superscriptaddress,showpacs,amsmath,amssymb,footinbib,longbibliography]{revtex4-2}

\usepackage[english]{babel}
\usepackage{amsmath,amssymb}
\usepackage{graphicx,xcolor}
\usepackage[unicode,bookmarksnumbered,breaklinks,colorlinks,linktocpage,
citecolor=blue,linkcolor=darkred,urlcolor=darkblue]{hyperref}
\usepackage{bm}
\usepackage{hyperref}

\definecolor{darkblue}{rgb}{0.0, 0.0, 0.55}
\definecolor{darkgreen}{rgb}{0.0, 0.2, 0.13}
\definecolor{darkred}{rgb}{0.55, 0.0, 0.0}

\usepackage{color}

\begin{document}

\title{Thermodynamics of the $S=1/2$ hyperkagome-lattice Heisenberg antiferromagnet}

\author{Taras Hutak}
\affiliation{Institute for Condensed Matter Physics,
    National Academy of Sciences of Ukraine,
    Svientsitskii Street 1, 79011 L'viv, Ukraine}

\author{Taras Krokhmalskii}
\affiliation{Institute for Condensed Matter Physics,
    National Academy of Sciences of Ukraine,
    Svientsitskii Street 1, 79011 L'viv, Ukraine}

\author{J\"{u}rgen Schnack}
\affiliation{Fakult\"{a}t f\"{u}r Physik, 
	Universit\"{a}t Bielefeld, 
	Postfach 100131, 33501 Bielefeld, Germany}

\author{Johannes Richter}
\affiliation{Institut f\"{u}r Physik, Otto-von-Guericke-Universit\"{a}t Magdeburg,
    P.O. Box 4120, 39016 Magdeburg, Germany}
\affiliation{Max-Planck-Institut f\"{u}r Physik komplexer Systeme,
	N\"{o}thnitzer Stra\ss e 38, 01187 Dresden, Germany}

\author{Oleg Derzhko}
\affiliation{Institute for Condensed Matter Physics,
	National Academy of Sciences of Ukraine,
	Svientsitskii Street 1, 79011 L'viv, Ukraine}
\affiliation{Fakult\"{a}t f\"{u}r Physik, 
	Universit\"{a}t Bielefeld, 
	Postfach 100131, 33501 Bielefeld, Germany}
\affiliation{Professor Ivan Vakarchuk Department for Theoretical Physics,
	Ivan Franko National University of L'viv,
	Drahomanov Street 12, 79005 L'viv, Ukraine}       

\date{\today}

\begin{abstract}
The $S=1/2$ hyperkagome-lattice Heisenberg antiferromagnet allows to study the interplay of geometrical frustration and quantum as well as thermal fluctuations in three dimensions. 
We use 16 terms of a high-temperature series expansion
complemented by the entropy-method interpolation
to examine the specific heat and the uniform susceptibility of this model.
We obtain thermodynamic quantities for several possible
scenarios determined by the behavior of the specific heat as $T\to 0$:
A power-law decay
with the exponent $\alpha=1,2$ and also $3$ (gapless energy spectrum) or an exponential decay (gapped energy spectrum). 
All scenarios give rise to a low-temperature peak in $c(T)$ (almost a shoulder for $\alpha=1$) at $T<0.05$, i.e., well below the main high-temperature peak.
The functional form of the uniform susceptibility $\chi(T)$ below about $T=0.5$ depends strongly not only on the chosen scenario but also on an input parameter $\chi_0\equiv\chi(T=0)$.   
An estimate for the ground-state energy $e_0$ depends on the adopted specific scenario but is expected to lie between $-0.441$ and $-0.435$.
In addition to the entropy-method interpolation we use the
finite-temperature Lanczos method to calculate $c(T)$ and $\chi(T)$ for finite
lattices of $N=24$ and $36$ sites.
A combined view on both methods leads us to favor the gapless scenario with $\alpha=2$ (but $\alpha=1$ cannot be excluded) and finite $\chi_0$ around $0.1$.
\end{abstract}

\maketitle

\section{Introduction}
\label{s1}

Frustrated quantum spin systems are a subject of intense 
ongoing research in the field of magnetism
\cite{highmagneticfields2002,quantummagnetism2004,frustratedspinsystems2005,Intro-Frust-Mag2011}.
Geometric frustration and quantum fluctuations may prevent 
any ground-state ordering even in three dimensions. 
Among several famous examples, the $S=1/2$ pyrochlore-lattice Heisenberg antiferromagnet has attracted much attention, 
being for decades a candidate for the realization of a spin-liquid state 
in three dimensions \cite{canals1998}.
After intense numerical studies,
a lattice symmetry breaking in the ground state has been revealed \cite{hagymasi2021,astrakhantsev2021,hering2022,schaefer2023}.

A closely related example is the $S=1/2$ hyperkagome-lattice Heisenberg antiferromagnet.
Inspired by experiments on 
the spinel oxide Na$_4$Ir$_3$O$_8$ \cite{okamoto2007},
in which low spin $d^5$ Ir$^{4+}$ ions reside on the vertices 
of a hyper\-kagome lattice,
several theoretical studies for the classical ($S\to\infty$) and quantum ($S=1/2$) Heisenberg antiferromagnet on such a lattice have been performed \cite{hopkinson2007,lawler2008b,zhitomirsky2008,zhou2008,lawler2008f,bergholtz2010,singh2012,wan2016,buessen2016}.
The main focus of these studies is at ground-state properties of the $S=1/2$ hyperkagome-lattice Heisenberg antiferromagnet. 
For the ground state of this model
a gapped quantum spin liquid with topological order \cite{lawler2008b} 
and 
a gapless quantum spin liquid with spinon Fermi surfaces \cite{lawler2008f} 
were proposed by Lawler {\it et al.}.
While the former proposal implies a nonzero spin gap, the latter one points to gapless spin excitations.
In contrast, Bergholtz {\it et al.} \cite{bergholtz2010} 
proposed a valence bond crystal with a 72 site unit cell as the ground state of this model
(thus, breaking translational symmetry but not spin-rotational one).
This proposal implies a spin gap with a huge number of singlet excitations below the lowest triplet state
and thus a power law for the specific heat and a vanishing susceptibility for vanishing temperature.

The finite-temperature properties of the $S=1/2$ hyperkagome-lattice Heisenberg antiferromagnet have also been considered \cite{lawler2008f,bergholtz2010,singh2012,buessen2016}.
For the gapless quantum spin liquid of Ref.~\citep{lawler2008f},
it was argued that $c(T)\propto T^{2}$ at low $T$ 
(similar to what is observed for Na$_4$Ir$_3$O$_8$ \cite{okamoto2007}) 
and that $\chi(T)$ is constant at low $T$ 
(again in agreement with experimental data for Na$_4$Ir$_3$O$_8$ \cite{okamoto2007}).
Application of the pseudofermion functional renormalization
group \citep{buessen2016} also shows that $\chi(T)$ exhibits no divergence down to zero temperature, but only a very weak increase.
In addition, high-temperature series expansions for $c$ and $\chi$ were developed
and compared with the experimental data for Na$_4$Ir$_3$O$_8$ \cite{singh2012}.

On the experimental side,
apart from the mentioned iridate compound Na$_4$Ir$_3$O$_8$ \citep{okamoto2007},
there are other candidates for a solid-state realization of the 
hyperkagome-lattice Heisenberg anitiferromagnet,
see, e.g., Refs.~\cite{koteswararao2014,chillal2020,sana2023}.
Note, however, that the $5d$-based transition-metal oxides, 
such as Na$_4$Ir$_3$O$_8$,
are known for having a large spin-orbit coupling 
so that the pure nearest-neighbor Heisenberg Hamiltonian apparently should 
be augmented by other terms relevant for such materials \cite{huang2017}.
Detailed comparisons of $c(T)$ and $\chi(T)$ between theory of Refs.~\cite{lawler2008f,singh2012,buessen2016} and experimental data for Na$_4$Ir$_3$O$_8$ \cite{okamoto2007} exhibit noticeable discreapancies roughly below $J/2$ ($J$ is about 300~K for Na$_4$Ir$_3$O$_8$) and even at higher temperatures for the specific heat.
The authors attributed this disagreement to an incomplete
subtraction of nonmagnetic contribution to the
experimentally measured $c(T)$ \cite{singh2012} and an insufficiency of the spin-isotropic Heisenberg model for description 
of the $S=1/2$ hyperkagome antiferromagnet Na$_4$Ir$_3$O$_8$ \cite{bergholtz2010,huang2017}.
Among various spin-anisotropic perturbations one may single out the Dzyaloshinskii-Moriya term, Kitaev term, and symmetric exchange anisotropic term \cite{huang2017}.
 
The aim of the present paper is to examine the
finite-temperature properties of the $S=1/2$ hyperkagome-lattice
Heisenberg antiferromagnet -- a benchmark model of frustrated
quantum magnets.
At the moment, we do not intend to compare theoretical findings with experiments since there are no good solid-state examples of such a model yet.
There are not so many methods applicable to tackle the thermodynamics of three-dimensional frustrated quantum spin systems. 
Quantum Monte Carlo suffers from the sign problem \cite{HeS:PRB00}, 
exact diagonalization or finite-temperature Lanczos methods 
are restricted to too small lattices \cite{JaP:PRB94,ScS:IRPC10,SRS:PRR20}, 
the density-matrix renormalization group technique requires 
a mapping via a ``snake'' path to a one-dimensional system \cite{USL:JMMM13}. 
Besides,
the pseudofermion functional renormalization group approach focuses on the wave-vector-dependent susceptibility  \cite{buessen2016},
whereas one more universal method, 
the rotation-invariant Green's function method
\cite{Richards1971,Kondo1972,Shimahara1991,Barabanov94,Ihle1997,mueller2018,mueller2019}, 
has not been applied to the $S=1/2$ hyperkagome-lattice
Heisenberg antiferromagnet so far.

In our study, we utilize the high-temperature series expansions to the order of $\beta^{16}$ ($\beta=1/T$)
provided by Singh and Oitmaa in Ref.~\cite{singh2012}.
Singh and Oitmaa used the high-temperature series to compute various thermodynamic properties 
down to a temperature
\footnote{Temperatures are given as multiples of the exchange interaction. $T=0.25$ thus means $T/J = 0.25$. In our study we set the nearest-neighbor Heisenberg antiferromagnetic interaction $J=1$.}
of about $T\approx 0.25$ \cite{singh2012}.
However, this range can be extended down to zero temperature
if one combines the series expansion with possible assumptions about the low-energy spectrum of the spin model within the framework of the so-called ``entropy method''.
The entropy-method interpolation of high-temperature series 
expansions was suggested by Bernu and Misguish \cite{bernu2001} 
and later used in several studies 
\cite{misguich2005,bernu2015,schmidt2017,bernu2020,derzhko2020,grison2020,gonzalez2022,hutak2023}.
In the present paper,
we unite the series expansion \cite{singh2012} and the entropy method \cite{bernu2001,misguich2005,bernu2015,bernu2020,derzhko2020,gonzalez2022}
to get the temperature dependence for the specific heat $c(T)$ and the uniform susceptibility $\chi(T)$ of the $S=1/2$ hyperkagome-lattice Heisenberg antiferromagnet over the full temperature range.
We also obtain a prediction for the ground-state energy of the model $e_0$,
which provides self-consistency of the entropy-method calculations.
Our entropy-method results are accompanied by finite-temperature Lanczos method data  
for finite lattices up to 36 sites.

The remainder of this paper is organized as follows. 
In Section~\ref{s2} we introduce the model 
and briefly explain the methods to be used for obtaining the thermodynamic quantities.
Then, in Section~\ref{s3}, we report our results for the ground-state energy $e_0$, the specific heat $c(T)$, and the uniform susceptibility $\chi(T)$.
Finally, we summarize our findings in Section~\ref{s4}.

\section{Model and methods}
\label{s2}

\begin{figure}
\includegraphics[width=0.995 \columnwidth]{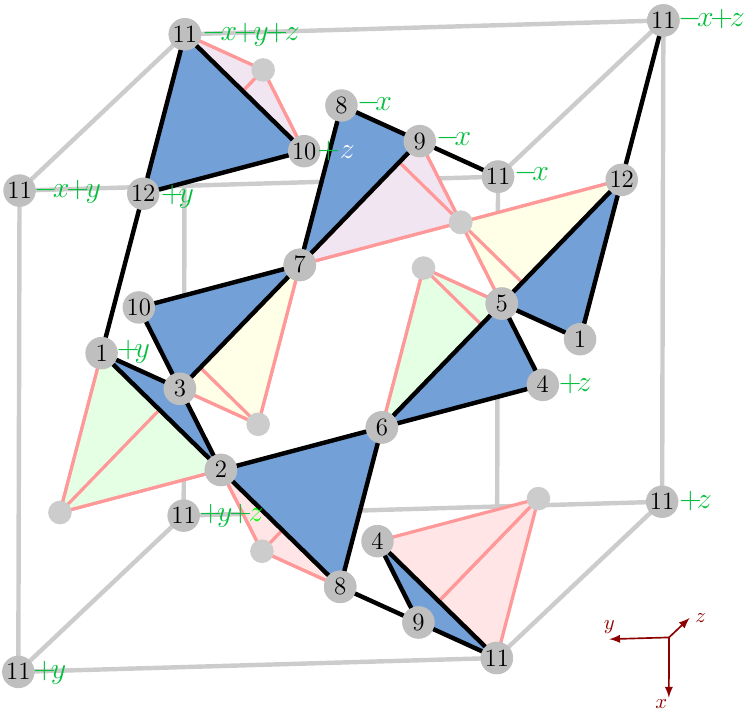}
\caption{The hyperkagome lattice.
The unit cell contains 12 equivalent sites ${\bm{r}}_{1},\ldots,{\bm{r}}_{12}$ (\ref{eq1}) denoted as $1,\ldots,12$.
Each site is connected to its four closest neighbors by bonds (black lines). For more details of the lattice see the main text.
We also display the underlying pyrochlore lattice.}
\label{rrfig1}
\end{figure}

The hyperkagome lattice has been described in several papers.
It can be viewed as a three-dimensional network of corner-sharing triangles with 12 sites in a cubic unit cell. 
It also can be viewed as a 1/4 depleted pyrochlore lattice, 
meaning that three out of the four sites in every tetrahedron are occupied by spins.
As a result, 
each spin of the three-dimensional hyperkagome lattice has only four nearest neighbors
just as for the two-dimensional kagome lattice. 
There are several different conventions regarding the coordinates of lattice sites
(see, e.g., Refs.~\cite{huang2017,jin2020,chern2021,pohle2023}).
According to Fig.~\ref{rrfig1},
we define the sites on the hyperkagome lattice sites by 
$\bm{R}_{\bm{n}\alpha}=\bm{R}_{\bm{n}}+\bm{r}_{\alpha}$.
Here,
$\bm{R}_{\bm{n}}=n_{x}\bm{e}_{x}+n_{y}\bm{e}_{y}+n_{z}\bm{e}_{z}$,
where 
$n_{x},n_{y},n_{z}$ are integers 
and 
$\bm{e}_{x}=(1,0,0)$, $\bm{e}_{y}=(0,1,0)$, $\bm{e}_{z}=(0,0,1)$,
generates a simple cubic lattice.
Moreover, 
the origins of the 12 equivalent sites in the unit cell may be defined by 
$\bm{r}_{\alpha}$, $\alpha=1,\dots,12$
with
\begin{eqnarray}
\label{eq1}
\bm{r}_{1}{=}\frac{1}{4}\left({-}2,0,2\right),
\bm{r}_{2}{=}\frac{1}{4}\left({-}1,3,2\right),
\bm{r}_{3}{=}\frac{1}{4}\left({-}2,3,1\right),
\nonumber\\
\bm{r}_{4}{=}\frac{1}{4}\left({-}1,1,0\right),
\bm{r}_{5}{=}\frac{1}{4}\left({-}2,1,3\right),
\bm{r}_{6}{=}\frac{1}{4}\left({-}1,2,3\right),
\nonumber\\
\bm{r}_{7}{=}\frac{1}{4}\left({-}3,2,1\right),
\bm{r}_{8}{=}\frac{1}{4}\left(0,2,2\right),
\bm{r}_{9}{=}\frac{1}{4}\left(0,1,1\right),
\nonumber\\
\bm{r}_{10}{=}\frac{1}{4}\left({-}3,3,0\right),
\bm{r}_{11}{=}\left(0,0,0\right),
\bm{r}_{12}{=}\frac{1}{4}\left({-}3,0,3\right).	
\end{eqnarray}
In Fig.~\ref{rrfig1},
we denote 
$\bm{r}_{1},\dots,\bm{r}_{12}$ by $1,\ldots,12$.
In addition,
we display
13 sites of the nearby unit cells by 
$11-x+y+z$, $11-x+z$, $8-x$ and so on,
where, e.g., $11-x+y+z$ means $\bm{r}_{11}-\bm{e}_{x}+\bm{e}_{y}+\bm{e}_{z}$, 
and so on.

In the present paper we consider the isotropic Heisenberg Hamiltonian on the hyperkagome lattice,
which is given by
\begin{equation}
\label{eq2}
H=\sum_{\langle\bm{m}\alpha;\bm{n}\beta\rangle}
\bm{S}_{\bm{m}\alpha}\cdot\bm{S}_{\bm{n}\beta}
\ .
\end{equation}
No extra interactions which may be relevant for solid-state compounds like Na$_4$Ir$_3$O$_8$ are included here,
i.e., we treat a kind of idealized minimal model without
complicated details but which already provides an interplay
between geometrical frustration and quantum and thermal
fluctuations. 
In Eq.~(\ref{eq2}),
the sum runs over the nearest-neighbor 
bonds of the hyperkagome lattice,
and $\bm{S}_{\bm{m}\alpha}$ represents the $S=1/2$ 
spin-vector operator 
at the lattice site $\bm{R}_{\bm{m}\alpha}$. 
Expanding the sum in Eq.~(\ref{eq2}) for fixed $\bm{m}$, 
one gets 24 bonds,
that is, 
15 bonds connecting the sites within the unit cell with the same cell index $\bm{m}$
and
9 bonds connecting the sites of the unit cell $\bm{m}$ 
with the sites of the neighboring unit cells 
$\bm{m}-\bm{e}_{x}$,
$\bm{m}-\bm{e}_{y}$,
$\bm{m}-\bm{e}_{z}$,
$\bm{m}-\bm{e}_{x}+\bm{e}_{y}$,
$\bm{m}+\bm{e}_{x}-\bm{e}_{z}$,
and 
$\bm{m}+\bm{e}_{y}-\bm{e}_{z}$,
see Fig.~\ref{rrfig1}.
The remaining 4 bonds in Fig.~\ref{rrfig1},
i.e., the ones which connect the sites 
$1+y$ and $12+y$,
$8-x$ and $9-x$,
$9-x$ and $11-x$,
$11-x+y+z$ and $12+y$
(cf. the bonds connecting the sites 
$1$ and $12$,
$8$ and $9$,
$9$ and $11$,
$11-x+z$ and $12$), 
are shown here for the sake of clarity.

It is worth noting that the hyperkagome lattice has similarities 
with the two-dimensional kagome lattice (corner-sharing triangles in two dimensions), 
as well as  with the three-dimensional pyrochlore lattice (corner-sharing  tetrahedrons in three dimensions), which could be considered the ``mother" crystal structure, see Fig.~\ref{rrfig1}.
An important property is that all three lattices support dispersionless (flat) one-magnon bands.
The shortest closed loop on the hyperkagome lattice beyond the triangles is a decagon; it involves ten bonds. 
The shortest cycle on the kagome and pyrochlore lattices beyond the triangles is a hexagon;
it involves six bonds.
Since the even regular polygon (decagon or hexagon) is surrounded by equilateral triangles, 
one expects a localized-magnon state, which lives on a decagon or hexagon, and belongs to a flat band, for more details see Refs.~\cite{derzhko2007,derzhko2015}.

In the remaining part of this section, we briefly explain the exploited methods:
Numerics for finite-size lattices and high-temperature series complemented with the entropy-method interpolation.
Here, we only report the key elements necessary to state our results in Sec.~\ref{s3}.

First, we determine numerically temperature dependencies for periodic lattices of $N=12$ sites (exact diagonalization) and $N=24,\,36$ sites (finite-temperature Lanczos method);
for a similar study of the $S=1/2$ pyrochlore-lattice Heisenberg antiferromagnet see Refs.~\cite{chandra2018,derzhko2020}.
Since the unit cell for the hyperkagome lattice contains 12 sites,
finite-lattice numerics is restricted to one unit cell \cite{lawler2008f} and two or three unit cells arranged as a chain.
Within the finite-temperature Lanczos method, the sum over an orthonormal basis in the partition function is replaced in a Monte-Carlo fashion by a much smaller sum over $R$ random vectors where each random vector is employed for a trace estimation \cite{JaP:PRB94}.
In the present study we take $R = 200$ for $N = 24$ and $R = 20$ for $N = 36$.
More details about finite-lattice calculations can be found in Refs.~\cite{JaP:PRB94,ScS:IRPC10,SRS:PRR20,spin:258,richter2010spin}.
Our numerical results for finite-size lattices are reported and discussed in Sec.~\ref{s3}.

Second, we utilize the high-temperature series expansion up to 16th order, 
which was reported in Ref.~\cite{singh2012}
(the Magdeburg HTE code \cite{schmidt2011,lohmann2014} yields the same series of the specific heat and the static uniform susceptibility, however, only up to 13th order),
and employ the entropy method \cite{bernu2001,misguich2005,bernu2015}
to obtain temperature dependencies at all temperatures for infinite lattice.

\begin{figure}
\includegraphics[width=0.995 \columnwidth]{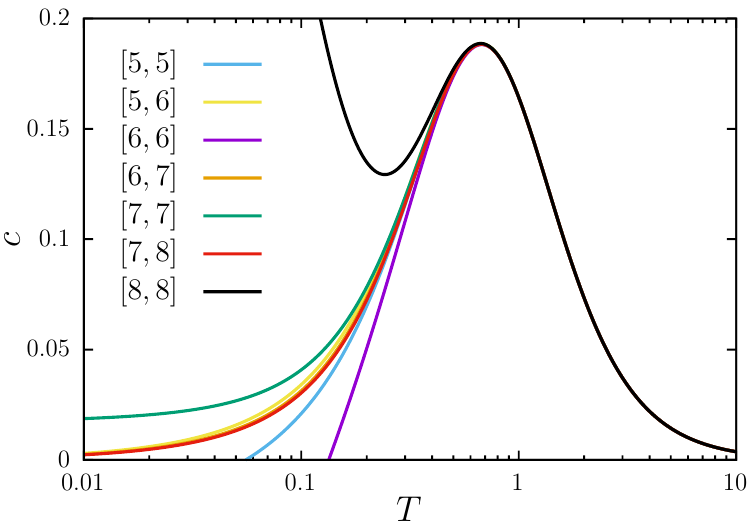}
\includegraphics[width=0.995 \columnwidth]{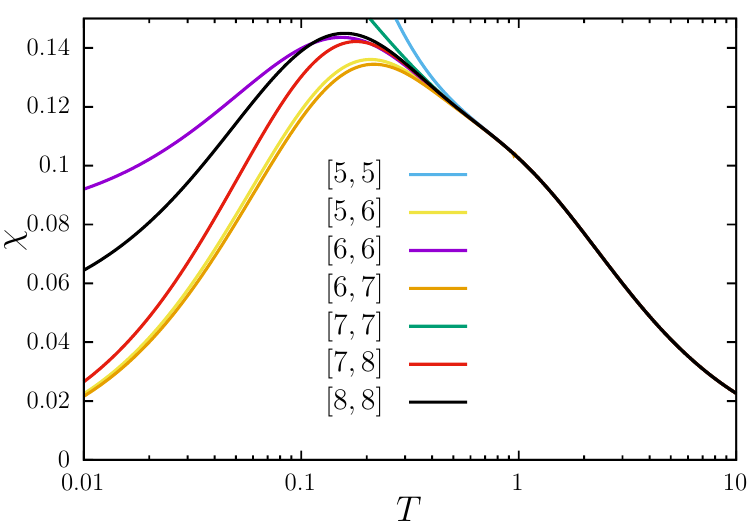}
\caption{Pad\'{e} approximants of the high-temperature series \cite{singh2012} for (top) the specific heat and (bottom) the uniform susceptibility.
They start to deviate from each other in both panels below $T\approx 0.5$.}
\label{rrfig2}
\end{figure}

As is well known,
the raw high-temperature series expansion may be improved by simple Pad\'{e} approximants $[u,d](T)=P_u(\beta)/Q_d(\beta)$.
Here, $P_u(\beta)$ and $Q_d(\beta)$ are polynomials of order $u$ and $d$, $u+d\le 16$, 
and the series expansion of  $[u,d](T)$   
coincides
with the high-temperature series of $c$ or $\chi$ up to $16$th order with respect to $\beta=1/T$.
Comparing close to diagonal Pad\'{e} approximants in Fig.~\ref{rrfig2},
we conclude that they start to deviate notably one from another 
below $T \approx 0.5$ and thus can reproduce the high-temperature peak 
of $c(T)$ at $T\approx 0.67$, but not any of the specific features of $\chi(T)$ 
since $\chi(T)$ increases monotonously to temperatures well below $T=0.5$ and also has got 
its maximum below $T=0.5$.

Within the entropy method \citep{bernu2001,misguich2005,bernu2015} one interpolates the entropy (per site) $s$ as a function of 
the mean (internal) energy (per site) $e$, $s(e)$. 
As $e$ approaches its maximal value 
$e_\infty=E(T\rightarrow\infty)/N=\text{tr}(H)/N=0$, 
the entropy is known from high-temperature series expansion, 
$s(e)=\ln2 +\sum_{i>1}a_ie^i$
(i.e., the coefficients $a_2,\ldots,a_{16}$ are known, see Ref.~\cite{bernu2001}).  
As $e$ approaches its minimal (ground-state) value $e_0$,
the entropy behaves as 
$s(e)\propto (e-e_0)^{\alpha/(1+\alpha)}$
if $c(T)=A T^\alpha$ for $T\to 0$ (gapless low-energy excitations)
or as
$s(e)\propto -[(e-e_0)/\Delta](\ln[\Delta(e-e_0)]-1)$
if $c(T)\propto e^{-\Delta/T}/T^2$ for $T\to 0$ (gapped low-energy excitations).
Next, we interpolate, instead of $s(e)$, an auxiliary function $G(e)$, different for the two types of low-energy excitations (the choice of $G(e)$ was discussed in detail in Refs.~\cite{bernu2001,misguich2005}), which immediately gives $s(e)$.
Such approximate quantities acquire the subscript ``app''.
For the gapless case we have
\begin{eqnarray}
\label{eq3}	
G(e){=}\frac{\left[s(e)\right]^{\frac{1+\alpha}{\alpha}}}{e-e_0} 
\to 
G_{\rm app}(e){=}\frac{\left(\ln 2\right)^{\frac{\alpha}{1+\alpha}}}{-e_0}
\frac{P_u(e)}{Q_d(e)}
\ ;
\nonumber\\
s_{\rm app}(e)=\left[\left(e-e_0\right)G_{\rm app}(e)\right]^{\frac{\alpha}{1+\alpha}}
\ .
\end{eqnarray}
And for the gapped case we have
\begin{eqnarray}
\label{eq4}
G(e){=}\left(e-e_0\right)\left[\frac{s(e)}{e-e_0}\right]^{\prime} 
\to 
G_{\rm app}(e){=}\frac{\ln 2}{e_0}\frac{P_u(e)}{Q_d(e)}
\ ;
\nonumber\\
\frac{s_{\rm app}(e)}{e-e_0}=\frac{\ln 2}{-e_0}-\int_{e}^0{\rm d}\xi\frac{G_{\rm app}(\xi)}{\xi-e_0}
\ .
\end{eqnarray}
Here,
$P_u(e)$ and $Q_d(e)$ are the polynomials of order $u$ and $d$, $u+d\le 16$,
and the series expansion of the quotient  $[u,d](e)=P_u(e)/Q_d(e)$ 
coincides 
with the Maclaurin series of $G(e)$ 
known up to $16$th order.
Besides,
the prime denotes the derivative with respect to $e$.
Knowing the dependence $s(e)$, we obtain the desired temperature dependence 
of the specific heat $c(T)$ in parametric form:
$T=1/s^{\prime}(e)$
and
$c=-[s^{\prime}(e)]^2/s^{\prime\prime}(e)$.
Finally,
we can calculate 
either the prefactor $A$,
$A_{\rm app}=[\alpha^{1+\alpha}/(1+\alpha)^\alpha][G_{\rm app}(e_0)]^\alpha$,
for the gapless case 
or 
the energy gap $\Delta$,
$\Delta_{\rm app}=-1/G_{\rm app}(e_0)$,
for the gapped case. 
In the presence of a (small) external magnetic field $h$ one gets the entropy $s_{\rm app}(e,h)$ which yields the uniform susceptibility $\chi$ via the relations: 
$m=[1/s^\prime(e,h)]\partial s(e,h)/\partial h$,
$\chi=m/h$ ($h\to 0$).
For further details see Refs.~\cite{bernu2001,misguich2005,bernu2015,bernu2020,derzhko2020,gonzalez2022}. 

Thus, to obtain the thermodynamic quantities within the framework of the entropy method 
one needs, 
besides the high-temperature series for $c$ and $\chi$, 
to know
i) the ground-state energy $e_0$,  
ii) how $c(T)$ vanishes as $T\to 0$,
and 
iii) the value of $\chi_0\equiv\chi(T=0)$ in the case of gapless low-energy excitations.  
Even if the precise value of $e_0$ is not available and both gapless and gapped excitations are acceptable, 
one can proceed as in Ref.~\cite{bernu2020}.
First, one has to assume some reasonable value $e_0$
in order to explore a certain region of $e_0$ systematically.
Second, one has to assume the exponent $\alpha$ in the case of a gapless spectrum
or one has to assume that the spectrum is gapped.
Then, for the assumed $e_0$ and gapless/gapped energy spectrum one has
to calculate within the entropy method the specific heat $c(T)$ 
using all $n_{\rm P}$ available Pad\'{e} approximants $[u,d](e)$.  
There are $n+1$ Pad\'{e} approximants based on the series up to $n$th order.
We discard from the very beginning four Pad\'{e} approximants $[n,0],\,[n-1,1],\,[1,n-1],\,[0,n]$ so that $n_{\rm P}=n-3$.
Next, one has to examine the ``closeness'' of all $n_{\rm P}$ profiles $c(T)$. 
To this end, we inspect them thoroughly from some high temperature $T_{\rm i}$ down to $T_{\rm f}<T_{\rm i}$ with temperature steps $\Delta T$.
If the absolute value of the difference of a certain $c$ from
the arithmetic mean value for this bundle, $\overline{c}$, at a running temperature $T$ ($T_{\rm f}\le T\le T_{\rm i}$) is less than, e.g., 0.001,
this $c$ belongs to the set of ``coinciding'' Pad\'{e} approximants. 
Otherwise, this Pad\'{e} approximant is discarded 
and not considered for lower temperatures.
According to Refs.~\cite{bernu2020,gonzalez2022},
a large number of coinciding curves $n_{\rm cP}$,
or more precisely a large value of $p=n_{\rm cP}/n_{\rm P}$,
provides evidence that the assumptions made about $e_0$ and 
the low-energy excitations are self-consistent.

\begin{figure}
\includegraphics[width=0.995 \columnwidth]{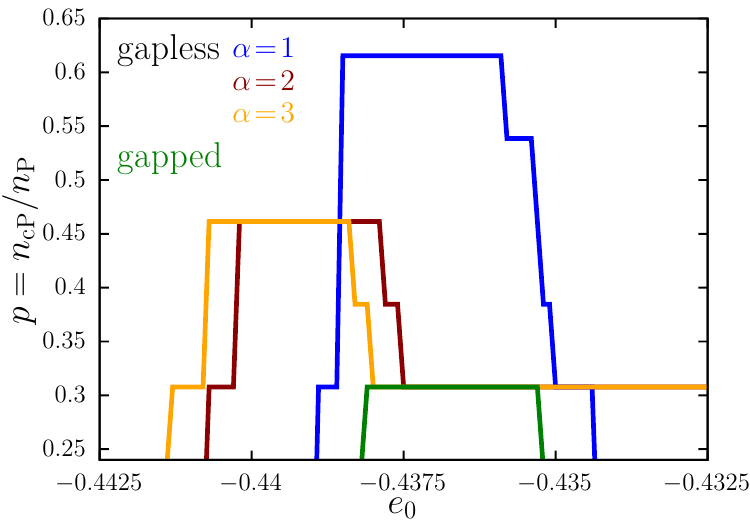}
\caption{The ratio of the number of ``coinciding'' entropy-method Pad\'{e} approximants $n_{\text{cP}}$ to the number of all considered entropy-method Pad\'{e} approximants $n_{\rm P}$, $p=n_{\text{cP}}/n_{\rm P}$, based on the series of 16th order
as a function of the chosen value of $e_0$.
Here $T_{\rm i}=0.5$, $\Delta T=0.025$, $T_{\rm f}=0.1$, see the main text.
We consider several assumptions, gapless and $\alpha=1$ (blue), $\alpha=2$ (red), $\alpha=3$ (orange) or gapped (green) low-energy excitations.}
\label{rrfig3}
\end{figure}

In Fig.~\ref{rrfig3} we illustrate such an analysis  
based on $n_{\rm P}=13$ Pad\'{e} approximants
(obtained from the 16th order high-temperature series expansion) in Eqs.~(\ref{eq3}) or (\ref{eq4}) for the specific heat $c(T)$ under the assumption of a gapless spectrum (blue, red, orange) or a gapped spectrum (green).
Here we set $T_{\rm i}=0.5$, $\Delta T=0.025$, $T_{\rm f}=0.1$.
If $e_0$ is taken in the range $[-0.438\,5, -0.435\,9]$
assuming a gapless spectrum and $\alpha=1$,
i.e., $c(T)=AT$ as $T\to 0$,  
we find that $n_{\text{cP}}=8$ and $p\approx 0.62$.
In addition, for the prefactor $A$ we
obtain the interval $[5.62,7.74]$.
Similarly,
if $e_0$ is taken in the range $[-0.440\,2, -0.437\,9]$ assuming $c(T)=AT^2$ as $T\to 0$, 
we find that $n_{\text{cP}}=6$, $p\approx 0.46$, whereas $A$ belongs to the interval $[493,727]$.
And for $e_0$ taken in the range $[-0.440\,7, -0.438\,4]$ under
the assumption $c(T)=AT^3$ as $T\to 0$,
we find $n_{\text{cP}}=6$ and $p\approx 0.46$.
Finally, assuming a gapped spectrum and taking $e_0$ in the range $[-0.438\,1, -0.435\,3]$, 
we find $p=4/13\approx0.31$.
In addition, the energy gap $\Delta$ for the lower
$e_0=-0.438\,1$ is $0.025$, whereas for the higher
$e_0=-0.435\,3$ it is $0.018$.
All these findings are illustrated in Fig.~\ref{rrfig3}.

Acting in accordance with the strategy of Refs.~\cite{bernu2020,gonzalez2022},
we may estimate the entropy-method prediction for the ground-state energy $e_0$ as follows:
Under the assumption of gapless excitations, 
$e_0$ depends on $\alpha$ but remains within the range $[-0.440\,7, -0.435\,9]$,  
whereas under the assumption of gapped excitations, 
$e_0\in [-0.438\,1, -0.435\,3]$.
In what follows, 
we use this missing input parameter $e_0$ 
for the entropy method, 
considering all assumptions about $c(T)$ as $T\to 0$,
as well as the minimal and maximal values of $e_0$ to obtain the shaded areas for $c(T)$ and $\chi(T)$, see Sec.~\ref{s32}.
We note in passing that the uniform susceptibility $\chi(T)$ is less convenient for 
seeking a large value of $p=n_{\text{cP}}/n_{\rm P}$,
since it requires the additional parameter $\chi_0$ if the spectrum is gapless.

More details about the entropy-method calculations can be found
in
Refs.~\cite{bernu2001,misguich2005,bernu2015,bernu2020,derzhko2020,gonzalez2022}.
Our entropy-method results are reported and discussed in Sec.~\ref{s3}.

\section{Results}
\label{s3}

\subsection{Ground-state energy $e_0$}
\label{s31}

We begin with the discussion of the ground-state energy of the
$S=1/2$ hyperkagome-lattice Heisenberg antiferromagnet.
Various proposals about the nature of the ground state, i.e., spin liquids or valence-bond crystals, 
yield $e_0=-0.424$ \citep{lawler2008f} or $e_0=-0.430\,115$ \citep{bergholtz2010}. 
Exact diagonalizations for $N=12,24,36$ yield $-0.453\,74$, $-0.446\,33$, $-0.445\,10$, 
that, apparently, are overestimated values of the thermodynamically large systems.
As explained above,
to provide consistency of the entropy-method calculations,
we have to assume for $e_0$ the values in the range $[-0.441, -0.435]$:
This is a combination of several possible scenarios of either a gapless (with $\alpha=1,2,3$) or a gapped energy spectrum. 
Yet another plausible simple approach to determine $e_0$ from the 
high-temperature series expansion 
\cite{hagymasi2021} yields $e_0$ about $-0.448$.
The determination of $e_0$ based on the high-temperature series expansion seems to be rather formal,
since it does not use any specific picture for the ground state of the model at hand.
However, the experience from other models, including exactly solvable ones and 
precisely examined numerically ones, gives hints that it may yield quite reasonable predictions 
\cite{bernu2020,gonzalez2022}.

It is worth noting that the ground-state energy for the kagome lattice is quite close: 
$-0.438\,6(5)$ \cite{yan2011,depenbrock2012}, 
$-0.438\,7$ \cite{laeuchli2019}, 
but for the pyrochlore lattice it is rather different: 
$-0.490(6)$ \cite{hagymasi2021}, 
$-0.483\,1(1)$ \cite{astrakhantsev2021},
$-0.489$ \cite{schaefer2023}. 

\subsection{Thermodynamic properties}
\label{s32}

\begin{figure}
\includegraphics[width=0.995 \columnwidth]{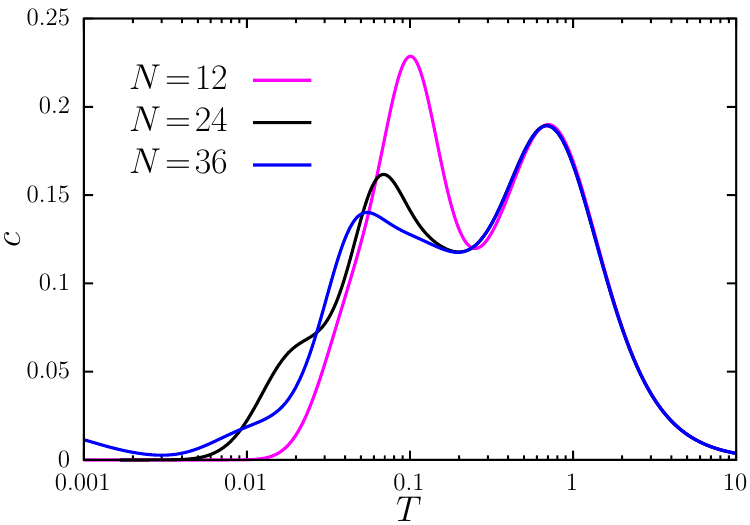}
\includegraphics[width=0.995 \columnwidth]{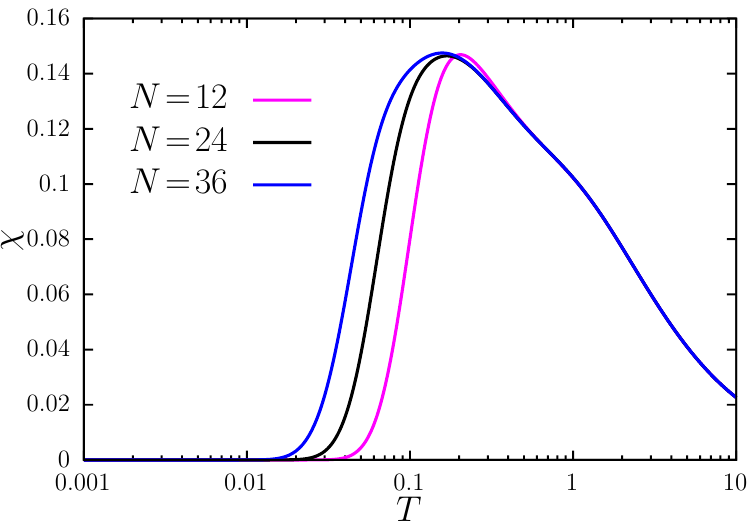}
\caption{Finite-lattice results for (top) the specific heat and (bottom) the uniform susceptibility of the $S=1/2$ hyperkagome-lattice Heisenberg antiferromagnet.
Exact-diagonalization ($N=12$) 
and 
finite-temperature Lanczos method 
($N=24$ and $N=36$) 
data.
The results for $N=24$ and $N=36$ differ from each other below about $T\approx 0.2$.}
\label{rrfig4}
\end{figure}

\begin{figure}
\includegraphics[width=0.995 \columnwidth]{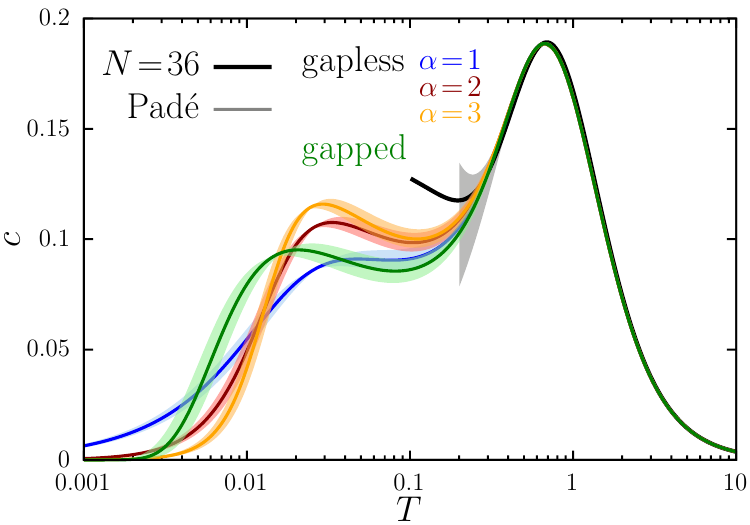}
\includegraphics[width=0.995 \columnwidth]{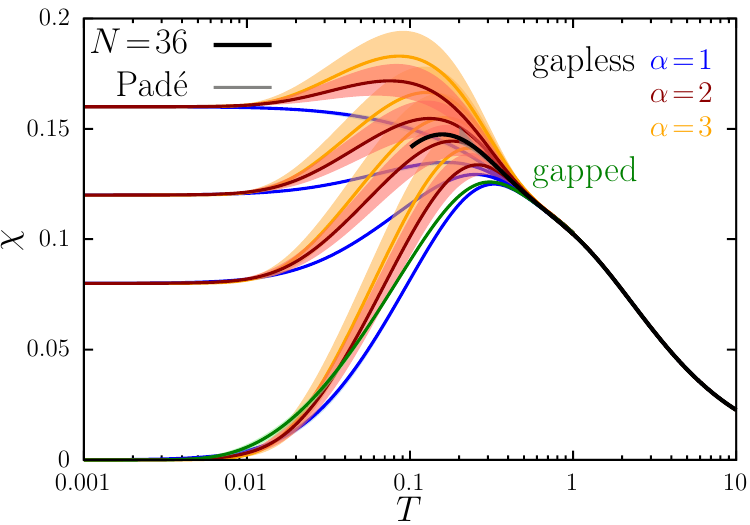}
\caption{Entropy-method results 
for (top) the specific heat and (bottom) the uniform susceptibility  
of the $S=1/2$ hyperkagome-lattice Heisenberg antiferromagnet. 
Blue, red, orange curves correspond to the gapless spectrum ($c=A T^{\alpha}$ as $T\to 0$) with $\alpha=1,2,3$, respectively, 
and 
green ones to the gapped spectrum ($c\!\propto\!e^{-\Delta/T}/T^2$ as $T\to 0$). 
The shaded area (light blue, light red, light orange, and light green) represents the region of $e_0$ 
where $p$ has a maximum (see Fig.~\ref{rrfig3}).
We also show $N=36$ data (black, $T\ge 0.1$)
and
two simple Pad\'{e} approximants $[7,7]$ and $[8,8]$ 
for $c(T)$ and $\chi(T)$  
and color in gray the region between them ($T\ge0.2$).
In the case of gapless excitations, we examine the four values of $\chi_0$: $0,\,0.08,\,0.12$, and $0.16$.}
\label{rrfig5}
\end{figure}

We pass to the finite-temperature properties of the $S=1/2$ hyperkagome-lattice Heisenberg antiferromagnet.
First, in Fig.~\ref{rrfig4} we report the temperature dependence of the specific heat $c(T)$ and the uniform susceptibility $\chi(T)$ obtained for finite lattices of $N=12,\,24,\,36$. 
Second, in Fig.~\ref{rrfig5} we report $c(T)$ and $\chi(T)$ obtained by the entropy method. 
Here, several possibilities, i.e., the gapless spectrum with $\alpha=1,2,3$ or the gapped spectrum, were considered, see blue, red, orange, or green curves, respectively.
The ground-state energy $e_0$ was determined from the analysis
of $c(T)$ as was explained in Sec.~\ref{s2}. 
We used $[8,8](e)$ in Eqs.~(\ref{eq3}) or (\ref{eq4}) 
as well as the region of $e_0$ where $p$ has a maximum, see Fig.~\ref{rrfig3}, in order to estimate the spread of the derived functions.
For the gapless excitations we set $\chi_0=0,\,0.08,\,0.12,\,0.16$. 

Let us now discuss the thermodynamic quantities of the $S = 1/2$ hyperkagome-lattice Heisenberg antiferromagnet in some detail. 
As it follows from the upper panel of Fig.~\ref{rrfig4}, the high-temperature peak of the specific heat does not show any finite-size scaling; it is already provided by the calculations for one unit cell ($N=12$). On these grounds, we thus speculate that the curve of the specific heat at temperatures of the high-temperature peak and above represents the thermodynamic
limit \footnote{In contrast, the results for the $S=1/2$ pyrochlore-lattice Heisenberg antiferromagnet of $N=32$ sites \cite{derzhko2020} reflect the thermodynamic limit only for $T>0.7$, well above the temperature of the high-temperature peak of $c(T)$. Therefore, the finite-lattice results for the hyperkagome case allow a reliable discussion of thermodynamic properties for much lower temperatures down to $T\approx 0.2$.}, see also $N=36$ data in Fig.~\ref{rrfig5}. The position of the low-temperature peak, on the other hand,
does depend on the size. 
Moreover, the height decreases notably with growing $N$.
The results of the entropy method in Fig.~\ref{rrfig5} refer to the infinite lattice. 
As it follows from the upper panel of Fig.~\ref{rrfig5},
the specific heat $c(T)$ besides the high-temperature peak at $T\approx 0.67$
has an additional low-temperature maximum, which is about two
times smaller and occurs below $T=0.05$ under both
assumptions of gapless and gapped excitations. For the case of
the gapless excitations with $\alpha=1$, the low-temperature
peak is so small that it is perceived as a shoulder, see the
blue curve in the upper panel of Fig.~\ref{rrfig5}. 

As can be seen in the lower panel of Fig.~\ref{rrfig4}, the maximum of $\chi(T)$ has a mild dependence on system size.
Moreover, the height remains practically unchanged.
This behavior can be traced back to the size of the
singlet-triplet gap for these systems.
Its value is $\Delta_{\rm s{-}t}\approx 0.383, 0.216, 0.136$ for $N=12, 24, 36$, respectively. 
According to the entropy-method analysis reported in the lower panel of Fig.~\ref{rrfig5}, 
the uniform susceptibility $\chi(T)$ behaves identically at $T$ above about $0.5$ for 
gapless and gapped excitations. For lower temperatures, 
the behavior of $\chi(T)$ depends on the adopted scenario and the $\chi_0$ value for gapless excitations.
In the case of gapless spin excitations, 
the uniform susceptibility approaches $\chi_0>0$ as $T\to0$ and
displays or does not display a maximum roughly below $T=0.3$
depending on the specific value $\chi_0$, cf. blue, red, orange
curves in the lower panel of Fig.~\ref{rrfig5}. 

An important general message that can be taken from
Fig.~\ref{rrfig5} is that the entropy-method and finite-system
numerical data (and even simple Pad\'{e} approximants for
$\chi$) favor the assumption of a gapless spectrum and a
quadratically vanishing specific heat at low temperature with
finite $\chi_0$ around $0.1$. However, a linear decay of the
specific heat cannot be excluded. 

It is worthy to put our results for the hyperkagome lattice in the context of prior work for the kagome and pyrochlore lattices. 
Concerning $c(T)$ (the upper panels of Figs.~\ref{rrfig4} and \ref{rrfig5}), 
its features at least at intermediate temperatures and above,
are quite similar to what is known for the kagome-lattice and also the square-kagome-lattice case 
(a peak at $T=0.67$, a shoulder of two times smaller height for $0.1<T<0.25$ \cite{SSR:PRB18,richter2022}),
but differ from those for the pyrochlore-lattice case, where 
only one peak in $c(T)$, but no additional low-temperature
feature such as peak or shoulder was found
\cite{derzhko2020,schaefer2020}.
Concerning $\chi(T)$ (the lower panels of Figs.~\ref{rrfig4} and \ref{rrfig5}),
it resembles the maximum of $\chi(T)$ for the finite-size kagome lattices \cite{SSR:PRB18} 
and for the infinite kagome lattice analyzed by the entropy method \cite{bernu2015}.
In contrast, for the pyrochlore lattice we have several
scenarios,
none of which can be excluded to date \cite{huang2016,derzhko2020,schaefer2020}. 
Thus, we may conclude that the three-dimensional hyperkagome
lattice is closer to highly frustrated two-dimensional lattices
(kagome, square-kagome) than to the three-dimensional pyrochlore
lattice.  
However, it is worth noting the difference: For the kagome
lattice the low-temperature peak of $c(T)$ moves to higher
temperatures with increasing $N$ \cite{SSR:PRB18}, opposite to
what is observed for the hyperkagome lattice (recall the top
panel of Fig.~\ref{rrfig4}). Thus, for the kagome lattice one
yields a low-temperature shoulder of the main peak in the
thermodynamic limit \cite{CRL:SB18}. 

\section{Summary and outlook}
\label{s4}

In the present paper, 
we used finite-lattice calculations and high-temperature series
expansions up to 16th order \cite{singh2012} complemented by
plausible assumptions about low-temperature properties
according to the entropy method
to obtain the specific heat $c(T)$ and the uniform
susceptibility $\chi(T)$ of the $S=1/2$ hyperka\-gome-lattice
Heisenberg antiferromagnet at all temperatures. 
Finite-lattice calculations are suitable to discuss the thermodynamics above $T=0.2$,
cf. $N=24$ and $N=36$ data in Fig.~\ref{rrfig4}.
The entropy method requires the knowledge of ground state and
low-energy excitations which is currently not
available. Therefore, we tested several scenarios (gapless
excitations with various exponents for a power-law decay of
$c(T\to0)$ and values of $\chi_0$ or gapped excitations) and
examined the coherence of the obtained results. 
Our main findings are as follows.
We observe a low-temperature peak in $c(T)$ (almost a shoulder
for $\alpha = 1$) at $T < 0.05$, i.e., well below the main
high-temperature peak.  
$\chi(T)$ below $T=0.5$ relies heavily on the adopted scenario.   
Combining finite-lattice and entropy-method results gives
evidence in favor of gapless spin excitations with a quadratic
low-temperature specific heat decay (although a linear one
cannot be ruled out)  
and finite $\chi$ at $T=0$ around $0.1$. 
Such results agree with the proposal of a gapless quantum spin liquid in Ref.~\cite{lawler2008f}
and the pseudofermion functional
renormalization group study of Ref.~\cite{buessen2016}. 
As a byproduct,
we have extracted the ground-state energy $e_0$ with a procedure explained in Sec.~\ref{s2}.
Although the value of $e_0$ depends on the adopted scenario, it was always restricted to  $[-0.441, -0.435]$.
We have found that the thermodynamics of the three-dimensional
hyperkagome-lattice Heisenberg antiferromagnet is quite similar
to the two-dimensional kagome-lattice one, but differs from that
on the pyrochlore lattice. 

Future work on the thermodynamics of the $S=1/2$ hyperkagome-lattice
Heisenberg antiferromagnet may be related to application of
specific tools to tackle the problem. 
For instance, this model represents a so-called flat-band
system: The one-magnon energy spectrum has a fourfold degenerate
dispersionless band, which is the lowest-energy band. The
flat-band states will be relevant at high fields and low
temperatures and their dominant contribution to thermodynamics
can be elaborated by special methods of flat-band systems, see
Refs.~\cite{derzhko2007,derzhko2015}. Such a program has been
realized for the $S=1/2$ kagome-lattice and pyrochlore-lattice
Heisenberg antiferromagnets in
Refs.~\cite{zhitomirsky2004,zhitomirsky2007},
and it might be applicable here, too.

\section*{Acknowledgements}

T.~H. was supported 
by the fellowship of the President of Ukraine for young scholars
and 
by the Projects of research works of young scientists of the National Academy of Sciences 
of Ukraine (project \# 29-04/18-2023, Frustrated quantum magnets at finite temperatures). 
O.~D. thanks J.~Stre\v{c}ka for the kind hospitality
at the MECO48 conference (Star\'{a} Lesn\'{a}, Slovakia, May 22-26, 2023).
O.~D. acknowledges the kind hospitality of the University of Bielefeld in October-December of 2023 
(supported by Erasmus+ and DFG).
This work was supported by the Deutsche Forschungsgemeinschaft (DFG SCHN 615/28-1 and RI 615/25-1).
T.~H., T.~K., and O.~D. acknowledge the support through the EURIZON project (Project No.~3025 ``Frustrated quantum spin models to explain the properties of magnets over wide temperature range''), which is funded by the European Union under Grant No.~871072.

\bibliography{hyperkag_refs_5,js-other,js-own}

\end{document}